\documentclass [12pt]{article}
\usepackage{amsmath,amssymb,cite}
\usepackage[english]{babel}
\usepackage[dvips]{graphicx}
\usepackage{indentfirst}
\numberwithin{equation}{section}
\begin{document}
\begin{center}
{\large{\bf Entanglement Entropy on Fuzzy Spaces}}\\
\bigskip
Djamel Dou$^{a,b}$, Badis  Ydri$^{c}$.

\bigskip
$^{a}${\it Institute of Exact Science and Technology, University
Center of Eloued, Eloued, Algeria}.\\

$^{b}${\it Dept of Physics and Astronomy, College of Science, King
Saud University, P.O. Box 2455 Riyadh 11451,
Saudi Arabia.}\\

$^{c}${\it Departement of Physics, Faculty of Sciences,
Badji Mokhtar-Annaba University, Annaba, Algeria .} \\
\end{center}
\begin{abstract}
We study the entanglement entropy of a scalar field in $2+1$
spacetime where  space is modeled by a fuzzy sphere and a fuzzy
disc. In both models we evaluate numerically the resulting
entropies and find that they are proportional to the number of
boundary degrees of freedom. In the Moyal plane limit of the fuzzy
disc the entanglement entropy per unite area (length) diverges if
the ignored region is of infinite size. The divergence is
(interpreted)
 of IR-UV mixing origin. In general we expect  the entanglement
entropy per unite area to be finite on a non-commutative space if
the ignored region is of finite size.
\end{abstract}

\section{Introduction }

It is believed that black hole thermodynamics hold important clues
to the nature  of the structure of spacetime at the very small
scale . For example the finiteness of black hole entropy is
understood as a direct manifestation of the discreetness  of
spacetime at the Planck scale, and points out to a necessary
reduction of the number of degrees of freedom on the horizon
\cite{sor,sh}. A piece of evidence for this comes from  the
contribution of the thermal atmosphere ( i.e the entropy stored in
the field quanta near the horizon ) to the black hole entropy
which was shown to be divergent in absence of any UV cutoff which
goes against the finiteness of black hole entropy \cite{t}. This
divergence is closely related to the entanglement entropy ( or
geometric entropy ) considered in \cite{s1,s} and  later in
\cite{sd}.

The notion of entanglement entropy originates from the simple
observation that an observer outside the horizon has no access to
the degrees of freedom behind the horizon. For this reason the
outside observer would describe the world with  a reduced density
matrix obtained by tracing out the unaccessible degrees of freedom
behind the horizon. If the exterior modes and the external modes
are correlated ``entangled'' the resulting density operator is
thermal even if the global state of the system were pure.

 The entanglement entropy has a UV divergence which can not be
renormalized away, at least when calculated  for fields in a fixed
background. Moreover in \cite{su} it was shown that the
divergences in the entanglement entropy are the same divergences
one must deal with when trying to renormalize the theory of
gravity coupled to matter, and therefore understanding the
finiteness of the entropy may not be possible without a complete
knowledge of the UV behavior of the theory.

However if we introduce a short distance  cutoff around the planck
length one obtains a finite entropy of the same order of magnitude
as that of a black hole. This entanglement entropy generally
scales like the area of the boundary. This has  led to many
speculations attributing the black hole entropy to the sum of all
entanglement entropies  of the fields in nature \cite{ted}.
Whether or not  the entanglement  of quantum  fields furnishes all
of the entropy or part of it, contribution of this type must be
present, and any consistent theory  must provide  for them in its
thermodynamic accounting.

Although the entanglement entropy has been extensively explored by
many authors most of the work was oriented toward computing it
using different regularization methods and studying the degree and
the universality of the divergences \cite{w,ks,imp}. For instance
in \cite{ks} it was shown that except in $1+1$-dimensions the
coefficient of the divergences is non-universal and depends on the
regularization method. Therefore any quantitative comparison of
different calculations was not possible.

The aim of this paper is to study the entanglement entropy on some
fuzzy spaces and draw possible conclusions about the entanglement
entropy on non-commutative spaces in general. The motivations for
considering the entanglement entropy on fuzzy spaces are many. One
obvious reason is the general hope that non-commutativity would
soften UV divergences and may render UV divergent quantities
finite as it is the case on fuzzy spaces, and at the same time
preserve the basic symmetry  of the space . For instance results
from finite-temperature non-commutative field theory showed that
the non-commutative model behaves as if it had many fewer degrees
of freedom in the UV than in conventional field theory. In some
cases the degrees of freedom were so drastically reduced that the
UV catastrophe could be avoided. See \cite{doug} and references
therein .

Another motivation is that non-commutative geometry arises
naturally in some limits of string theory in connection with
D-branes. In effect it has recently been suggested by many authors that
the microstates counting of non extremal black holes using field
theory dual  of string theory could be interpreted as arising from
entanglement \cite{st1,st2,st3}. In considering entanglement
entropy on fuzzy or noncommutative spaces one faces the question
of defining the boundary with respect to which the entanglement is
to be computed. Indeed  in view of the non locality of non
commutative theory one would expect any boundary to be fuzzy. This
also brings in the question how to define disjoint regions in
non-commutative or fuzzy spaces in order to properly define the
entanglement entropy. We will show in this paper that once the
field variables are properly chosen there is almost a unique way
to obtain entanglement between different degrees of freedom in
different regions.

This paper is organized as follows. In section $2$ we give a brief
review of the formalism which we will use. In particular we
compute the entanglement entropy for a free scalar theory defined
on the continuum sphere. In section $3$, part $1$ we compute the
entanglement entropy resulting from tracing half of a fuzzy sphere
and show that the result is proportional to the number of boundary
degrees of freedom or equivalently to the area of the boundary .
In part $2$ of section $3$ we reconsider the same problem on a
truncated Moyal plane, i.e a fuzzy disc, and obtain similar
results as in fuzzy sphere case. We also discuss the Moyal plane
limit and observe that if the ignored region is blown up a UV-IR
mixing phenomenon takes place. We conclude in section $4$ by
general discussions and possible improvement of our numerical
results.

\section{The Regularized theory}
In this section we review the main formalism we will be using and
 apply it to a regularized continuum sphere. For detail of the
formalism we refer the reader to \cite{s,sd}. For a review see
\cite{jap}. The Hamiltonians that we will consider in this paper
are of the standard form
 \begin{equation}\label{1}
    H=\frac{1}{2}\sum_{A,B} (\delta_{A,B}
    \pi^A\pi^B+V_{AB}\varphi^A\varphi^B)
\end{equation}
$V_{AB}$ is a real symmetric matrix with positive definite
eigenvalues. The case where $V$  has zero eigenvalues needs
special treatment. It corresponds to the case of a massless field
and we will return to it later. The normalized ground state of
$(\ref{1})$ is given in the Schrodinger representation by
\begin{equation}\label{2}
    <{\varphi^A}\mid 0> = \big[\det \frac{W}{\pi}\big]^{1/4}\exp \big
[-\frac{1}{2}W_{AB}\varphi^A\varphi^B\big]
\end{equation}
Where $W$ is the square root of the matrix $V$. The corresponding
density matrix is
\begin{equation}\label{3}
    \rho(\varphi,\varphi')= \big[\det \frac{W}{\pi}\big]^{1/2}\exp\big
[-\frac{1}{2}W_{AB}(\varphi^A\varphi^B+\varphi'^A\varphi'^B)\big]
\end{equation}
Now  if we consider  the information  on the fields degrees of
freedom $ \varphi^\alpha, \alpha=\overline{1,n}$ as unavailable,
we form a reduced density matrix by integrating over the  $
\varphi^\alpha, \alpha=\overline{1,n}$,
\begin{equation}\label{4}
   \rho_{\rm
red}({\varphi}^{n+1},{\varphi}^{n+2},...,{\varphi}^{'n+1},{\varphi}^{'n+2},...)=
    \int \prod_{\alpha=1}^{n} d{\varphi}^\alpha \rho(\varphi,{\varphi}')
\end{equation}
 The entanglement entropy is the  associated Von Newman entropy  of
$ \rho_{\rm red}$  defined by $ S= -{Tr} \rho_{\rm
red}\log\rho_{\rm red}$. The entanglement entropy for any
Hamiltonian of the form (\ref{1}) can be shown to be given by
\begin{equation}\label{5}
    S_{\rm ent}= \sum_i\bigg[
\log\big(\frac{1}{2}\sqrt{{\lambda}_i}\big)+
    \sqrt{1+{\lambda}_i}\log\bigg(\frac{1}{\sqrt{{\lambda}_i}}+
    \sqrt{1+\frac{1}{{\lambda}_i}}\bigg)\bigg]
\end{equation}
Where ${\lambda}_i$ are  the eigenvalues of the following matrix
\begin{equation}\label{6}
    \Lambda_{i,j}= -\sum_{\alpha=1}^{n}W^{-1}_{i\alpha}W_{\alpha j}
\end{equation}
$W_{\alpha j}$ and $ W^{-1}_{i\alpha}$ are  elements of $W$ and
$W^{-1}$ respectively  with $i,j$ running from $n+1$ to $N$ and
$\alpha$ from $1$ to $n$ . $\Lambda$ is an $(N-n)\times(N-n)$
matrix and $i,j$ run now from $n+1$ to $N$.

Before considering theories defined on fuzzy spaces we start with
a real free scalar field defined on spacetime with topology  $
\bf{R}\times \bf{S}^2$ where $\bf{R}$ is time and $\bf{S}^2$ the
spatial slice.The lagrangian is given by
\begin{eqnarray}
L=\frac{1}{2}\int d{\Omega}
\big(\dot{\phi}^2+{\phi}\big({\Delta}-\mu^2\big){\phi}\big).
\end{eqnarray}
For computational reasons it turns out that the theory is easily
regularized in the cylindrical coordinates. The Laplacian
${\Delta}$ in this coordinates is given by
\begin{eqnarray}
{\Delta}=\frac{1}{R^2}\frac{\partial}{{\partial}z}\big((R^2-z^2)
\frac{\partial}{{\partial}z}\big)+\frac{1}{R^2-z^2}\frac{{\partial}^2}{{\partial
}{\phi}^2}.
\end{eqnarray}
Expanding the field in Fourier modes by writing
\begin{eqnarray}
\phi=\frac{1}{\sqrt{2\pi}}\sum_{m=-\infty}^{+\infty}{\phi}_m(z)e^{-im\varphi}~,~
{\phi}_m^{*}(z)={\phi}_{-m}(z).
\end{eqnarray}
We find that the Lagrangian $L$ is the direct sum of individual
microscopic Lagrangians $L_m$ each associated with an allowed
value of the azimuthal  number $m$, viz

\begin{equation}\label{7}
    L=\sum_{m=-\infty}^{+\infty}L_m
\end{equation}
where

\begin{eqnarray}
L_m&=&\frac{1}{2}\int_{-R}^{R}dz\bigg[\dot{Q}_m^2-(1-\frac{z^2}{R^2})\big(
\frac{{\partial}Q_m}{{\partial}z}\big)^2-\big(\frac{m^2}{R^2-z^2}+{\mu}^2\big)Q_
m^2\bigg]
\end{eqnarray}

where $Q_0=\phi(z)$, $Q_m=\sqrt{2}{Re}\phi_m$ for $m>0$ and
$Q_m=\sqrt{2}{Im}\phi_m$ for $m<0$ .

 We regularize this model as follows.
The $z-$axis is replaced by a one-dimensional lattice, i.e
$z{\longrightarrow}z_n=n a$ where
$a=\frac{R}{N}{\longrightarrow}0$ is the lattice spacing and
$n=-(N-1),..,(N-1)$. The regularized microscopic Lagarangian for
one sector is given by

\begin{equation}\label{8}
    L_m= \frac{1}{2a}\sum_{n=-(N-1)}^{N-1} \bigg[(a
    \dot{Q}_{m,n})^2-(1-\frac{n^2}{N^2})\big(Q_{m,n}-Q_{m,n-1}\big)^2-(\frac{m^2}{N^2-n^2}+a^2\mu^2)Q_{m,n}^2\bigg]
\end{equation}
 The corresponding Hamiltonian is
\begin{equation}\label{9}
    H_m= \frac{1}{2a}\sum_{n=-(N-1)}^{N-1} \bigg[
    \pi_{m,n}^2+(1-\frac{n^2}{N^2})\big(Q_{m,n}-Q_{m,n-1}\big)^2+(\frac{m^2}{N^2-n^2}+a^2\mu^2)Q_{m,n}^2\bigg].
\end{equation}
By scaling the fields and shifting the summation variable the
Hamiltonian $H_m$ can be brought to the general form of equation
(\ref{1}), namely
\begin{equation}\label{10}
    H_m =\sum_{A,B=1}^{2N-1}\bigg[\delta_{A,B} \pi^A \pi^B+
    V^{(m)}_{AB}Q_m^A Q_m^B\bigg]
\end{equation}
 where
 \begin{eqnarray}
V_{AB}^{(m)}&=&\big(2-\frac{(N-A)^2}{N^2}-\frac{(N-A+1)^2}{N^2}+a^2{\mu}^2+\frac
{m^2}{N^2-(N-A)^2}\big){\delta}_{A,B}\nonumber\\
&-&(1-\frac{(N-A)^2}{N^2}){\delta}_{B,A-1}
-(1-\frac{(N-B)^{2}}{N^2}){\delta}_{A,B-1}
\end{eqnarray}

The macroscopic Hamiltonian is given by
\begin{equation}\label{11}
    H=\sum_{m=-\infty}^{+\infty}H_m
\end{equation}
If we now consider the degrees of freedom residing in the upper
hemisphere  unaccessible we construct the reduced density operator
for the ground state by integrating all the modes $ Q_m^{\alpha}$
for $\alpha= N,...,2N-1$ corresponding to positive $z$ for all
values of $m$.  According to  (\ref{11})   the resulting reduced
density operator is of the form
$$
\bigotimes_{m=-\infty}^{\infty}{\rho}_{\rm red}^{(m)}
$$
with ${\rho}_{\rm red}^{(m)}={\rho}_{\rm red}^{(-m)}$, the total
entanglement entropy is therefore given by
\begin{equation}\label{S_N}
    S_N= S_0+2\sum_{m=1}^{\infty} S_m
\end{equation}
 $S_m$ is computed by applying equation $(\ref{5})$ to the Hamiltonian
$H_m$
 with
\begin{equation}\label{12}
    \Lambda^{(m)}_{i,j}= -\sum_{\alpha=N+1}^{2N-1}W^{-1}_{i\alpha}W_{\alpha
j}
\end{equation}
The indices $i$ and $j$ run in the available region  i.e
$i,j=1,...,N$.
 For a given $N$ we compute numerically the entropy for different
 masses. This entails the computation of the eigenvalues of
 $\Lambda$ for each pair of values $(N,m)$ and  then  computing
the~ entropy  by  means of equation (\ref{S_N}). However to obtain
a
 numerical result the sum over $m$ must be cutoff at some value
$m_{max}$.
 This can be decided by the following simple observation similar to
that in \cite{sd} . For large $m$, $ m\gg N$, the $m$-dependent
term dominates over the other terms
 and $S_m$ can be computed perturbatively. It is found that for
 $m\gg N$
 $$
\Lambda^{(m)}_{a,b}= \delta_{a,N}\delta_{b,N}\frac{N^4}{4m^4}+
O((\frac{N}{m})^6)
$$
which gives for $S_m$
\begin{equation}\label{13}
    S_m\approx\frac{N^4}{16m^4}(1-log \frac{N^4}{16m^4})
\end{equation}
This  demonstrates that the sum over $m$ will converge and also
can be used to set an upper bound on the remaining of the sum
which can easily be seen to be negligible for $m\gg N^{4/3}$. The
large $m$ behavior of $S_m$ is very similar to that obtained in
$3+1$ by Srednicki \cite{sd}.

\begin{figure}[h]
\begin{center}
\includegraphics[width=9cm,angle=0]{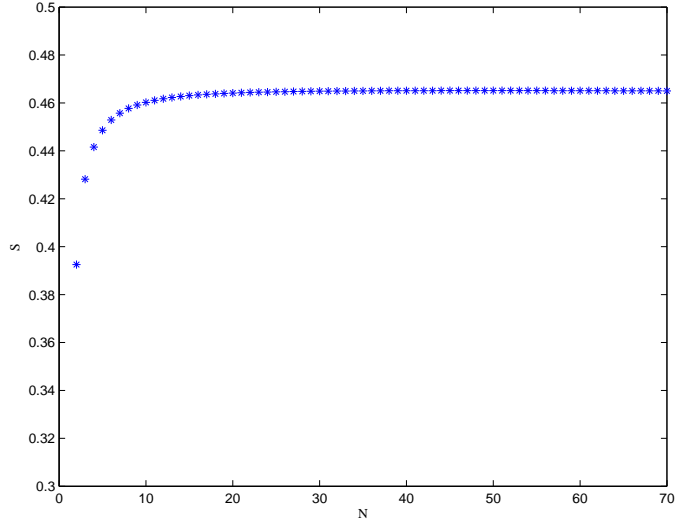}
\caption{{ The scaled entropy $S$ for a massless scalar field on
continuum sphere, with $m_{max}=2000$.}}
\end{center}
\end{figure}

The result for the scaled entropy  $ S=S_N/N$, $m_{\rm max} =2000$
and massless field is depicted on figure 1. It is observed that
the entropy converges for large $N$ to the value $0.465 N$. Hence
the resulting entropy is proportional to the area (circumference)
of the boundary with a proportionality coefficient which goes like
1/cutoff, via
\begin{equation}\label{}
    S_N= \frac{0.465}{2\pi a} A
\end{equation}
where $A =2\pi R$, which is the area  law in $2+1$ dimension.

\section{Fuzzy  spaces}

\subsection{The fuzzy sphere}
We start with a scalar field on ${\bf R}{\times}{\bf S}^2_N$ where
${\bf S^2}_N$  is a fuzzy sphere of  matrix dimension $N=2l+1$.
The action reads

\begin{eqnarray}
S_N=\frac{1}{N}\int dt L~,~L=\frac{1}{2}Tr \bigg(\dot{\phi}^2
+\phi\big[{\cal L}_i^2-{\mu}^2\big]\phi \bigg).\label{2.1}
\end{eqnarray}
The scalar field $\phi$ is an $N{\times}N$ hermitian matrix with
mass parameter $\mu$. The Laplacian ${\cal L}_i^2$ is the $ SU(2)$
Casimir operator given by ${\cal L}_i^2={\cal L}_1^2+{\cal
L}_2^2+{\cal L}_3^2$ with action defined by ${\cal
L}_i({\phi})=[L_i,\phi]$ and ${\cal
L}_i^2({\phi})=[L_i,[L_i,\phi]]$. The $L_i$ satisfy
$[{L}_i,{L}_j]=i{\epsilon}_{ijk}{L}_k$ and they generate the
$SU(2)$ irreducible representation of spin $l=\frac{N-1}{2}$.

First we observe that
\begin{eqnarray}
Tr{\phi}{\cal L}_i^2{\phi}=2{\phi}_{ab}M_{ab,cd}{\phi}_{cd}
\end{eqnarray}
where
\begin{eqnarray}
M_{ab,cd}&=&\bigg((L_i^2)_{bc}\delta_{da}-(L_i)_{bc}(L_i)_{da}\bigg)=\bigg(c_2{\delta}_{bc}\delta_{da}-(L_i)_{bc}(L_i)_{da}\bigg).
\end{eqnarray}
$c_2$ is the Casimir of the spin $l$ IRR of $SU(2)$, i.e
$c_2=l(l+1)$. Now we introduce two  {\it real} matrix scalar
fields by splitting $\phi$ as follows $\phi={\phi}_1+i{\phi}_2$.
The hermiticity of $\phi$ implies that ${\phi}_1^T={\phi}_1$ and
${\phi}_2^T=-{\phi}_2$, in other words ${\phi}_1$ is a real
symmetric $N{\times}N$ matrix whereas ${\phi}_2$ is a real
antisymmetric $N{\times}N$ matrix. By using also the fact that the
matrix $M_{ab,cd}$ satisfies $M_{ba,dc}=M_{ab,cd}$ it is a
straightforward calculation to show that

\begin{eqnarray}
Tr{\phi}{\cal
L}_i^2{\phi}=2({\phi}_1+{\phi}_2)_{ab}M_{ab,dc}({\phi}_1+{\phi}_2)_{cd}=2
\Phi_{ab}M_{ab,dc}\Phi_{cd}.
\end{eqnarray}
where $\Phi \equiv {\phi}_1+{\phi}_2$.  For explicit calculation
we use  the matrix form of $ L_1$ and $L_2$ in the base where
$L_3$ is diagonal. They are given  for arbitrary $l$ by
\begin{eqnarray}
(L_1)_{ab}=\frac{1}{2}[B_b\delta_{a,b+1}+B_a\delta_{a,b-1}]~,~
 (L_2)_{ab}=\frac{i}{2}[B_b\delta_{a,b+1}-B_a\delta_{a,b-1}]~,~(L_3)_{ab}=
A_a \delta_{a,b}
 \end{eqnarray}
 Where $ B_a = \sqrt{a(N-a)}$ and $ A_a =-a+\frac{N+1}{2}$. The indices
$a,b$ run from $1$ to $N=2l+1$. We can immediately compute
\begin{eqnarray}
Tr{\phi}{\cal
L}_i^2{\phi}=2{\Phi}_{ab}\big(c_2-A_aA_b\big){\Phi}_{ab}-{\Phi}_{ab}\big(B_{a-1}
B_{b-1}\big){\Phi}_{a-1,b-1}-{\Phi}_{ab}\big(B_aB_b\big){\Phi}_{a+1,b+1}.
\end{eqnarray}
Now, this expression suggests the introduction of the following
fields : $  Q^{(m)}$ defined by $Q^{(m)}_{a}={\Phi}_{a,a+m}$ and
$Q^{(-m)}_{a}={\Phi}_{a+m,m}$ for $m=0,..N-1$. More explicitly we
have for $m=0$ the field $Q^{(0)}$ with $N$ degrees of freedom
given by
\begin{eqnarray}
 Q^{(0)}=(
   \Phi^{11}, \Phi^{22},\cdot  \cdot \cdot ,\Phi^{NN}
).
\end{eqnarray}
For $m=+1$ we have  the fields $Q^{(+1)}$ and $Q^{(-1)}$ each with
$N-1$ degrees of freedom given by
\begin{eqnarray}
Q^{(+1)}=(
   \Phi^{12}, \Phi^{23}, \Phi^{34},  \cdot \cdot ,\Phi^{N-1,N})
~,~{Q}^{(-1)}=(
   \Phi^{21}, \Phi^{32}, \Phi^{43},  \cdot \cdot ,\Phi^{N,N-1})
\end{eqnarray}
For general positive $m$ we have the fields $Q^{(+m)}$ and
$Q^{(-m)}$ which contain each $(N-m)$ degrees of freedom given by

\begin{eqnarray}
Q^{(m)}=(
   \Phi^{11+m}, \Phi^{22+m}, \cdot  \cdot \cdot ,\Phi^{N-m,N}
)~,~Q^{(-m)}=(
   \Phi^{1+m,1}, \Phi^{2+m,2}, \cdot  \cdot \cdot ,\Phi^{N,N-m}
)
\end{eqnarray}
The last two fields $Q^{(N-1)}$ and $Q^{-(N-1)}$ contain one
degrees of freedom each, viz
\begin{eqnarray}
Q^{(N-1)}=(
   \Phi^{1N})~,~Q^{(-N+1)}=(
   \Phi^{N1}).
\end{eqnarray}
Using this parametrization we can show that
\begin{eqnarray}
Tr{\phi}{\cal
L}_i^2{\phi}&=&2\sum_{m=-(N-1)}^{N-1}\sum_{a=1}^{N-|m|}Q_a^{(m)}\bigg[\big(c_2-A
_aA_{a+|m|}\big){\delta}_{a,b}
-\frac{1}{2}B_{a-1}B_{a-1+|m|}{\delta}_{a-1,b}-\frac{1}{2}B_{a}B_{a+|m|}{\delta}
_{a+1,b}\bigg]Q_{b}^{(m)}.\nonumber\\
\end{eqnarray}
Similarly we can compute
\begin{eqnarray}
Tr\big(\dot{\phi}^2-{\mu}^2{\phi}^2\big)=\sum_{m=-(N-1)}^{N-1}\sum_{a=1}^{N-|m|}
Q_a^{(m)}\big(-{\partial}_t^2+{\mu}^2\big)Q_{a}^{(m)}.
\end{eqnarray}
Hence the Hamiltonian  $H$ of the free theory takes the form
\begin{eqnarray}
H=\sum_{m=-(N-1)}^{N-1}
H_m=\sum_{m=-(N-1)}^{N-1}\sum_{a,b=1}^{N-|m|}\bigg[\frac{1}{2}(\pi^{(m)}_a)^2+\frac{1}{2}V_{ab}^{(m)}Q^{(m)}_a
Q^{(m)}_b\bigg].\label{2.2}
\end{eqnarray}
where
\begin{eqnarray}
V_{ab}^{(m)}=2\bigg[\big(c_2+\frac{{\mu}^2}{2}-A_aA_{a+|m|}\big){\delta}_{a,b}
-\frac{1}{2}B_{a-1}B_{a-1+|m|}{\delta}_{a-1,b}-\frac{1}{2}B_{a}B_{a+|m|}{\delta}
_{a+1,b}\bigg].\label{V}
\end{eqnarray}
and $\pi^{(m)}_a = \dot{Q}^{(m)}_a$ . With this result one can see
that the free theory splits into $2(2l)+1$ independent sectors $
\{\mathcal{H}_m\} , m=-(N-1),\cdot\cdot\cdot\cdot, (N-1)$, each
sector $\mathcal{H}_m$ has $N-|m|$ degrees of freedom ($N-|m|$
coupled harmonic oscillator) and described by a Hamiltonian $H_m$.
The ground state density matrix is easily  seen to be
\begin{eqnarray}
\rho=\bigotimes_{m=-(N-1)}^{N-1} \rho^{(m)}.
\end{eqnarray}
It is worth mentioning here  that the formalism of the fuzzy
sphere bears a lot of similarity to the lattice regularization of
the continuum discussed in the previous section. However the fuzzy
sphere provides a natural cutoff for the quantum number $m$ which
is playing a similar role to the azimuthal  number of the
continuum sphere.

Having brought the Hamiltonian  on the fuzzy sphere to the form we
want we are ready now to discuss the entanglement entropy in the
fuzzy sphere setting. First we note that in order to introduce the
entanglement entropy we need to divide the field degrees of
freedom into two sets residing in two disjoint regions
corresponding, say, to the upper and lower fuzzy hemispheres. In
order to do that rigourously one needs to give precise criterion
that allows one to decide whether a given set of two fields ( two
matrices) have a disjoint support on the fuzzy sphere. Although we
believe that deriving such criterion may not be difficult we shall
content ourselves to some heuristic and intuitive arguments. We
first note that the form of the Hamiltonian suggests strongly that
we take each sector ${\cal H}_m $ and trace over half of the
degrees of freedom. For a fixed $N$ and $m$ the number of degrees
of freedom in the sector ${\cal H}_m $ is $N-|m|$, if $N-|m|$ is
even we trace out the following degrees of freedom
\begin{eqnarray}
Q^{(m)}_{1},Q^{(m)}_{2},\cdot\cdot\cdot,Q^{(m)}_{k},
k=\frac{N-|m|}{2}
\end{eqnarray}
if $N-|m|$ is odd we have two options, either we trace out
\begin{eqnarray}
Q^{(m)}_{1},Q^{(m)}_{2},\cdot\cdot\cdot,Q^{(m)}_{k}, k=
\frac{N-|m|-1}{2}
\end{eqnarray}
or we trace out
\begin{eqnarray}
Q^{(m)}_{1},Q^{(m)}_{2},\cdot\cdot\cdot, ,Q^{(m)}_{k},k=
\frac{N-|m|+1}{2}
\end{eqnarray}
However both options lead to the same entanglement entropy for
large $N$ and the degrees of freedom $
Q^{(m)}_{\frac{N-|m|+1}{2}}$ will be interpreted as boundary
degrees of freedom and there are $N$ of them.
 This corresponds in the original matrix notation
to dividing the matrix $\phi$ into two parts, upper left triangle
$ \phi_{U}$ and right lower triangle $\phi_{L}$ . For  example for
the first option above for $N=5$ the $\phi_{U}$ and $\phi_{L}$
will look as follows

\begin{eqnarray}
\phi_{U}=\left (\begin{array}{ccccc}
   {\phi}_{11} & {\phi}_{12} & {\phi}_{13} & {\phi}_{14} & 0 \\
   {\phi}_{21} & {\phi}_{22} & {\phi}_{23} & 0 & 0 \\
   {\phi}_{31} & {\phi}_{32} & 0 & 0 & 0 \\
   {\phi}_{41} & 0 & 0 & 0 & 0 \\
   0 & 0 & 0 & 0 & 0
 \end{array}\right)~,~
\phi_{L}=\left (\begin{array}{ccccc}
  0& 0 & 0 & 0 & \phi_{15}\\
    0 &0 & 0 & \phi_{24}  &  \phi_{25}\\
      0 & 0 & \phi_{33}  & \phi_{34} & \phi_{35}  \\
        0 &  \phi_{42} & \phi_{43} & \phi_{44}  & \phi_{45} \\
      \phi_{51}& \phi_{52} & \phi_{53} & \phi_{54} &  \phi_{55}
\end{array}\right)
\end{eqnarray}
The components $ \phi_{51}, \phi_{42}, \phi_{33},
\phi_{24},\phi_{15}$ are the boundary degrees of freedom.
$\phi_{U}$ and $\phi_{L}$ can be given the interpretation of
corresponding to functions with disjoint supports, one on the
lower half and the other on the upper half of the fuzzy sphere. A
hint for this comes from the observation that
${Tr}(\phi_{U}\phi_{L})=0$, for arbitrary $\phi_{U}$ and
$\phi_{L}$ and the fact that we are working in the basis where
$L_3$ is diagonal and therefore one can talk about negative and
positive $z$ coordinates ( i.e upper and lower hemisphere). Indeed
this is the only choice that leads to a non-zero entanglement
entropy. Now from equation $(\ref{2.2})$  one can easily see that
the resulting reduced density operator will be
\begin{equation}\label{2.3}
{\rho}_{\rm red} = \bigotimes_{m =-(N-1)}^{N-1} {\rho}_{\rm
red}^{(m)}
\end{equation}
and the associated entropy  is
\begin{equation}\label{2.4}
    S_l= S_0 +2 \sum_{m=1}^{2l}S_m , ~~ N=2l+1.
\end{equation}
where $S_m$ is computed using equation $(\ref{5})$ with  $
\Lambda^{(m)}$ defined by $\Lambda^{(m)}_{ij}=
-\sum_{\alpha=1}^{k}W^{-1}_{i\alpha}W_{\alpha j}$ where $W$ is the
square root of the potential matrix $V^{(m)}$ given by equation
(\ref{V}) . The indices $i,j$ run from $k+1$ to $N-m$ (the
available region).
 Since we expect our entropy to be proportional
to the circumference of the equator which can be said to be a
fuzzy circle, we define  a scaled entropy by dividing the entropy
by the square root of the Casimir $c_2$, viz
\begin{equation}\label{}
    S= \frac{S_N}{\sqrt{l(l+1)}}
\end{equation}
  We compute the scaled entropy $S$ numerically  for
  $N=4\cdot\cdot\cdot\cdot\cdot,200$ and run several numerical
  calculations with different values of the mass. The results are
  depicted on figure 2 . We can immediately see that the different
  curves with  masses $\mu^2 = 0,10^{-5},1^{-3},1$ approach  the same
value for large
  $N$. For example for $l=600$ the scaled entropy for the different
masses is
  equal to $0.39$ and the discrepancy  between the four different
  values being in the third digit.

   We can thus safely extrapolate to very large $l$ or $N$ and
   conclude that the entropy is given by
  \begin{equation}\label{2.6}
    S_N= 0.39 \sqrt{l(l+1)}.
\end{equation}
Since the radius of the fuzzy sphere  can be scaled with $l$
keeping the noncommutativity parameter $ \theta=
\frac{R}{\sqrt{l(l+1)}}$ fixed, we can write the entropy in the
following form
\begin{equation}\label{2.7}
    S_l= 0.39 \frac{R}{\theta}= \frac{0.39 A}{2\pi\theta}
\end{equation}
This is exactly the area law in two dimension.

Some remarks about our numerical evaluations for the entropy are
in order. As in any numerical calculation, numerical errors are
inevitable. In our calculation numerical errors originate from
rounding errors and iteration, however it turns out that they are
really tiny, in worst cases of order $10^{-10}$. Indeed none of our matrices is ill-conditioned or nearly singular to a
working precision, therefore one should not for instance expect
significant errors to occur by inversion.

Now, despite the fact that numerical errors are really small the
exact numerical value  for the leading contribution to the scaled
entropy can not be extrapolated from our numerical
result, because as figure $2$ shows there are
subleading corrections which  are nonvanishing, although small,
for finite $l$ and affect the rate of convergence.

 Although one
can not obtain the form of the subleading corrections, our
numerical calculations show that they are negligible for large
$l$, for example they are of order $10^{-3}$ for $l=600$ and for  the values of the mass we used.

\begin{figure}[h]
\begin{center}
\includegraphics[width=12cm,angle=0]{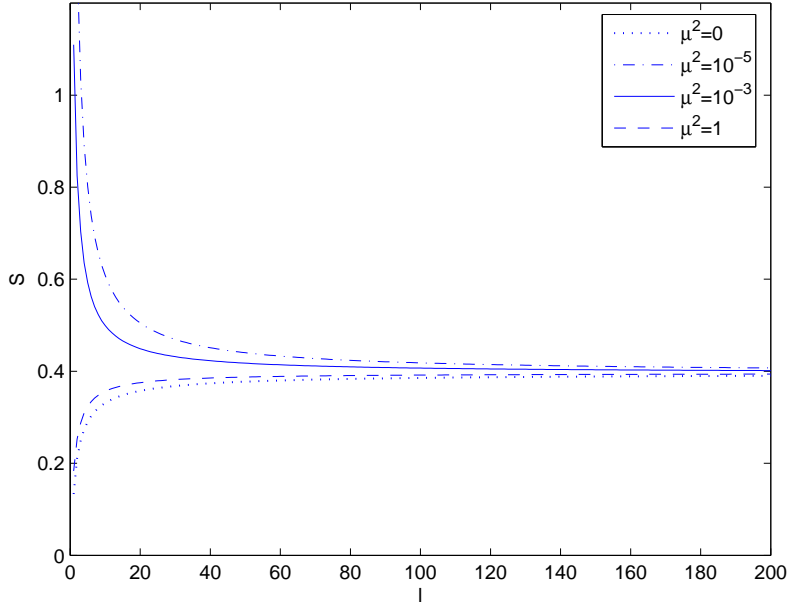}
\caption{{ The scaled entropy $S$ resulting from tracing out half
of the fuzzy sphere.}}
\end{center}
\end{figure}

Having established the area law, let us now turn to a simpler and
more interesting interpretation of the result obtained. Since
equation $(\ref{2.6})$  is valid  for large $l$ we can write it as
\begin{equation}\label{2.9}
  S_l= 0.19(2l+1) \sim \log 2^{2l+1}
\end{equation}
Therefore the entropy is directly proportional to the
 number of the degrees of freedom on the dividing boundary, in this
case being a ``fuzzy circle''. This result fits with the picture
of the horizon (simulated here by the equator) as being divided or
quantized into  small cells  of Planck size with each cell
carrying roughly one bit of information \cite{sor}. The picture
geos along the same line with old suggestion that the area of a
black hole have a quantized spectrum. See for example
\cite{brian}.

We conclude  this section by few comments about the result for the
massless case. For relatively small values of $N$ or $l$ there is
a clear mass dependence which dominates the behavior of the
entropy. The entropy is a decreasing function of the mass as can
be seen from figure $2$. However it shows a discontinuity at zero
mass where it drops from large values for small masses to small
values for zero mass. However this discontinuous behavior
disappears for very large $l$ and the entropy has a smooth
dependence on the mass. To understand the zero mass discontinuity
we go back to the original potential. It is easy to see from
equation (\ref{V}) that the potential $ V^{(0)}$ of the sector
$\mathcal{H}_0$ is a singular matrix in the massless case and therefore one
 can not apply directly  the formalism of section $2$. However, once treated carefully
one can easily show that the sector $\mathcal{H}_0$ gives zero
contribution to the entanglement entropy and therefore must be
projected out from the sum in equation  (\ref{2.4}).  In the
massless case  the associated entropy  is therefore
\begin{equation}\label{2.10}
    S_N= 2 \sum_{m=1}^{N-1}S_m
\end{equation}
All the terms $S_m$ are computed using the same formalism of
section $2$ since none of the potentials $V^m$ is singular.
Moreover, for large $l$ (the continuum limit) the main
contribution to $S$ comes from the sectors with $m$ different from
zero, and the term $S_0$ becomes irrelevant which explains the
smooth behavior  for large $l$. For instance already at $l=400$
and for $\mu^2=0.1$ both equations (\ref{2.10}) and (\ref{2.4})
give for $S_N$  the same value $S= 0.39$ . Now the reason for
which previous lattice computation, like the one presented in the
first section and the one in \cite{sd}, were not sensitive to this
effect is due to the fact that those lattice regularizations break
the translation symmetry or rotation symmetry to which the zero
modes of the potential are related.

\subsection{Moyal plane and the fuzzy disc}
We consider now a scalar theory on ${R}{\times}{R}_{\theta}^2$
where ${ R}_{\theta}^2$ is now the Moyal plane. The action is
given by :
\begin{equation}
 S=\frac{1}{2}\int dt Tr ( \dot{\phi}^{2}+\phi\big(
\nabla^{2}-{\mu}^2)\phi).\label{disc}
\end{equation}
The trace is infinite dimensional and the Laplacian is given in
terms of creation and annihilation operators $a$ and $a^{+}$ by
the experssion
\begin{equation}
\nabla^{2}\phi:=
-\frac{4}{\theta^{2}}[a^{+},[a,\phi]]=-\frac{4}{\theta^{2}}[a,[a^{+},\phi]].
\end{equation}
Let us recall that $[a,a^{+}]={\theta}$ where $\theta$ is the
noncommutativity parameter. The fuzzy disc is obtained from the
plane following \cite{liz} as follows. We consider finite
dimensional $N{\times}N$ matrices ${\phi}$, viz
\begin{equation}
\phi=\sum_{n,m=0}^{N-1}\phi_{mn}\mid
m><n\mid~,~{\phi}^{+}={\phi}~,~{\phi}_{nm}^{*}={\phi}_{mn}.
\end{equation}
Then it can be shown that the Laplacian $\nabla^{2}$ acts on a
 finite  dimensional space of dimension $(N+1)^2$, i.e $\nabla^{2}\phi$
is an$(N+1){\times}(N+1)$ matrix. The action on ${\bf
R}{\times}{\bf D}_N^2$ is thus given by (\ref{disc}) where the
trace $Tr$ is simply cut-off at $N$. We denote this trace by
$Tr_{N}$. The radius of the disc is given by
\begin{eqnarray}
R^2=N{\theta}.
\end{eqnarray}
In other words when $N{\longrightarrow}{\infty}$ we must take
${\theta}{\longrightarrow}0$ to recover a commutative disc.

For the purpose of computing the entanglement entropy on this
space the relevant piece of the action is given by

\begin{eqnarray}
V&=& -Tr_{N} \phi\big(
\nabla^{2}-{\mu}^2){\phi}=({\mu}^2+\frac{4}{\theta})Tr_{N}{\phi}^2+\frac{8}{{\theta}^2}Tr_{N}{\phi}^2a^{+}a-\frac{8}{{\theta}^2}Tr_{N}{\phi}a{\phi}a^{+}.
\end{eqnarray}
Explicitly we can compute
\begin{eqnarray}
\frac{\theta}{2}V&=&2\sum_{n,m=0}^{N-1}\bigg[(1+\frac{{\mu}^2\theta}{4}+n+m){\phi}_{nm}{\phi}_{mn}-\sqrt{nm}\phi_{nm}\phi_{m-1n-1}+\sqrt{(n+1)(m+1)}\phi_{nm}\phi_{m+1n+1}\bigg]\nonumber\\
\end{eqnarray}
or equivalently ( with $\tilde{\phi}_{nm}={\phi}_{n-1m-1}$ )
\begin{eqnarray}
\frac{\theta}{2}V&=&2\sum_{n,m=1}^{N}\bigg[(-1+\frac{{\mu}^2\theta}{4}+n+m)\tilde{\phi}_{nm}\tilde{\phi}_{mn}-\sqrt{(n-1)(m-1)}\tilde\phi_{nm}\tilde\phi_{m-1n-1}+\sqrt{nm}\tilde\phi_{nm}\tilde\phi_{m+1n+1}\bigg].\nonumber\\
\end{eqnarray}
By using the same trick we used on the fuzzy sphere; namely we
split the fields into symmetric part
$\phi^{(1)}_{mn}=\phi^{(1)}_{nm}$ and antisymmetric part
$\phi^{(2)}_{mn}=-\phi^{(2)}_{nm}$ by writing
$(\tilde{\phi})_{mn}=\phi^{(1)}_{mn}+i\phi^{(2)}_{mn}$, then we
recombine into the real field
$\Phi_{mn}=\phi^{(1)}_{mn}+\phi^{(2)}_{mn}$, we can put the above
action into the form
\begin{eqnarray}
-\frac{{\theta}}{2}Tr_{N} \phi\big(
\nabla^{2}-{\mu}^2){\phi}&=&2\sum_{n,m=1}^N{\Phi}_{nm}\big(-1+\frac{{\mu}^2\theta}{4}+A_n+A_m\big){\Phi}_{nm}\nonumber\\
&-&\sum_{n,m=1}^N\bigg[{\Phi}_{nm}\big(B_{n-1}B_{n-1}\big){\Phi}_{n-1m-1}+{\Phi}_{nm}\big(B_nB_m\big){\Phi}_{n+1m+1}\bigg]
\end{eqnarray}
where $A_n$ and $B_n$ are now defined by  $A_n=n$ and
$B_n=\sqrt{2n}$. The off diagonal elements have exactly the same
structure as the off diagonal elements on the fuzzy sphere,
whereas the diagonal elements here involve the sum $A_n+A_m$ as
opposed to the product $-A_nA_m$ on ${\bf S}^2_N$. Following the
same steps we have taken in the fuzzy sphere case we can write the
above action in the form

\begin{eqnarray}
-\frac{{\theta}}{2}Tr_{N} \phi\big(
\nabla^{2}-{\mu}^2){\phi}=\sum_{m=-(N-1)}^{N-1}\sum_{a,b=1}^{N-|m|}V_{ab}^{(m)}Q_a^{(m)}Q_{b}^{(m)}
\end{eqnarray}
where
\begin{eqnarray}
V_{ab}^{(m)}&=&2\bigg[\big(-1+\frac{{\mu}^2{\theta}^2}{4}+A_a+A_{a+|m|}\big){\delta}_{a,b}
-\frac{1}{2}B_{a-1}B_{a-1+|m|}{\delta}_{a-1,b}-\frac{1}{2}B_{a}B_{a+|m|}{\delta}_{a+1,b}\bigg]\nonumber\\
&=&2\bigg[\big(2a+|m|-1+\frac{{\mu}^2{\theta}}{4}\big){\delta}_{a,b}
-\sqrt{(a-1)(a-1+|m|)}{\delta}_{a-1,b}-\sqrt{a(a+|m|)}{\delta}_{a+1,b}\bigg].\nonumber\\
\end{eqnarray}
The fields $Q_a^{(m)}$ and $Q_a^{(-m)}$ ( $m {\geq}0$ ) are
defined in the same way as on ${\bf S}^2_N$, in other words
\begin{eqnarray}
&&2Q_a^{(m)}=2{\Phi}_{a,a+m}=(1-i){\phi}_{a-1,a+m-1}+(1+i){\phi}^{*}_{a-1,a+m-1}\nonumber\\
&&2Q_a^{(-m)}=2{\Phi}_{a+m,a}=(1-i){\phi}_{a+m-1,a-1}+(1+i){\phi}^{*}_{a+m-1,a-1}.
\label{ew}
\end{eqnarray}
Altogether we obtain $2N-1$ independent sectors  $ {\mathcal H}_m$
,$m=-N+1,-N+2,... ,N-2,N-1$.

For the fuzzy disc we shall consider  two different entanglement
entropies. The first one will  result from tracing out a smaller
sub-disc, and the second one from tracing out half of the fuzzy
disc. In the first case the ignored region will remain finite once
we consider the Moyal plane limit, while  in the second case the
ignored region blows up. Consider now the following $(n+1)\times
(n+1)$ sub-matrix with $n <N$
\begin{eqnarray}
\left (\begin{array}{ccccc}
 \phi _{00} & \phi_{01} & \phi_{02} & ... &\phi _{0n} \\
  \phi_{10} &\phi _{11} & \phi_{12} & ...  &  ...\\
  \phi_{20} & \phi_{21} &...  & ... & ...\\
  ... &  ... & ... & ...  & ...\\
  \phi_{n0} & ... & ... & ... &  \phi _{nn}
\end{array}\right).
\end{eqnarray}
The degrees of freedom in this matrix have support on a sub-disc
${\bf D}_n$. We are going to assume that this region is
unaccessible to outside observer and therefore we are going to
trace the degrees of freedom residing in this region. One can
verify that the sectors $\mathcal{H}_{\pm m}$ will not be
concerned by this operation if $|m|>n$. The resulting entanglement
will receive contribution only from the sectors $\mathcal{H}_{m}$
with $|m|\leq n$ and will be a function of $N$ and $n$. This can
be seen from equation (\ref{ew}) as follows. The first row
${\phi}_{0a}$ of the above matrix $\phi$ corresponds to
$Q_1^{(m)}$ with $a=m$ and $0{\leq}m{\leq}n$. The second row
${\phi}_{1a}$ corresponds to $Q_2^{(m)}$ with $a=m+1$ and
$-1{\leq}m{\leq}n-1$. The $n$th row ${\phi}_{n-1a}$ corresponds to
$Q_n^{(m)}$ with $a=m+n-1$ and $-(n-1){\leq}m{\leq}1$ while the
last row ${\phi}_{na}$ corresponds to $Q_{n+1}^{(m)}$ with $a=m+n$
and $-n{\leq}m{\leq}0$. Thus the unavailable degrees of freedom
are $Q_a^{(m)}$ with $-n{\leq}m{\leq}n$ and $1{\leq}a{\leq}n+1$.
Remark that for any fixed $m{\geq}0$ the index $a$ runs over the
range $1{\leq}a{\leq}n+1-m$.For $m\le 0$ the index $a$ runs over
the range $1-m{\leq}a{\leq}n+1$.

The resulting reduced density operator is therefore

\begin{equation}\label{}
{\rho}_{\rm red} = \bigotimes_{m =-n}^{n} {\rho}_{\rm red}^{(m)}
\end{equation}
and the corresponding entropy is given by  ( also with
${\rho}_{\rm red}^{(m)}={\rho}_{\rm red}^{(-m)}$ )

\begin{eqnarray}
S_N(n)= S_{0}+2\sum_{m=1}^{n} S_m
\end{eqnarray}
where $S_m$ is computed using the matrix $\Lambda^{(m)}$ given by
\begin{eqnarray}
\Lambda^{(m)}_{ij}=-\sum_{\alpha=1}^{n+1-m}W^{-1(m)}_{i\alpha}W^{(m)}_{\alpha
j}.
\end{eqnarray}
The available indices $i,j$ run from $n+1-m$ to $n$. Again we
define a scaled entropy $S$ by
\begin{eqnarray}
S=\frac{S_N(n)}{2n+1}.
\end{eqnarray}
 For a fixed $N$ we compute $S$ numerically for $n=
2\cdot\cdot\cdot\cdot,N-1$. The results for $N=400$ are depicted
on Figure $3$. For  all values of $n$ with  $30 \leq n \leq 370$
the scaled entropy is almost a constant being  $S=0.235$ for
$n=75$ and drops very slowly until it reaches $S=0.230$ for
$n=350$. The fact that $S$ can not be a constant is due  to the
finiteness of ratio of $n$ to $N$, which is related to the fact
that the entropy can not be a constant all the way to $n=N$, since
at this point we will have traced out all the disc and the result
must be zero. When $n$ approaches $N$ i.e $n>380$ the entropy
starts to decrease faster until it reaches zero. Therefore for
$N\gg n \gg 1$\footnote{The condition $N \gg n$ may not be needed
to obtain this result, although it is hard to decide by numerical
calculation how close $n$ can be to $N$.}, we can safely write

\begin{equation}\label{3.1}
  S_N(n)= 0.23(2n+1)
\end{equation}
\begin{figure}[h]
\begin{center}
\includegraphics[width=9cm,angle=0]{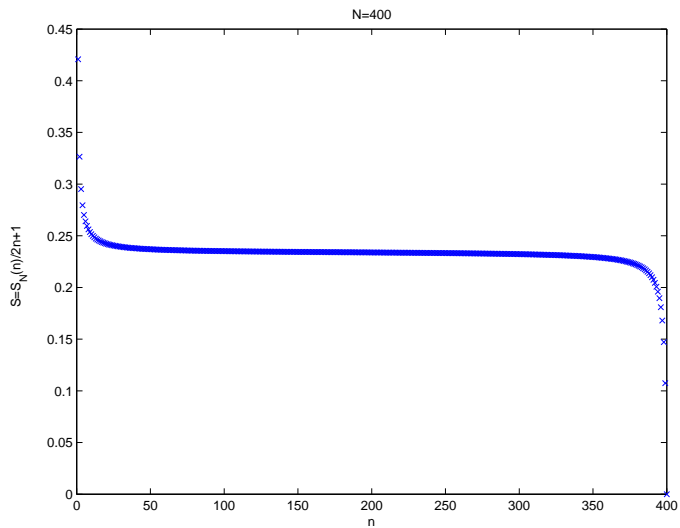}
\caption{{ The scaled entropy $S$ for a massless scalar resulting
from tracing out a sub-disc of a fuzzy disc .}}
\end{center}
\end{figure}
 Here again we find the entropy to be directly proportional to
the number of degrees of freedom on the boundary, and as can be
easily seen there  are $2n+1$ of them. It is interesting to
 note that  the leading contribution to $S_N(n)$  is independent
 of $N$.

Let us now consider the fuzzy disc and instead of tracing out a
smaller sub-disc consider the entropy resulting from tracing out
half of the disc (upper or lower half). The procedure is exactly
similar to the fuzzy sphere case. We define a scaled entropy $S=
S_N/{N}$ and compute numerically the resulting entropy for a
massless scalar for each $N$. The results are depicted on Figure
$4$. It is seen that $S$ is rapidly converging  towards a constant
value independent of $N$ equal to $0.341$, namely
\begin{equation}\label{3.2}
  S_N= 0.34N.
\end{equation}
In view of our previous discussions this result is also expected.
The entanglement entropy is proportional to the number of boundary
degrees of freedom which is $N+1$ in this case.

\begin{figure}[h]
\begin{center}
\includegraphics[width=9cm,angle=0]{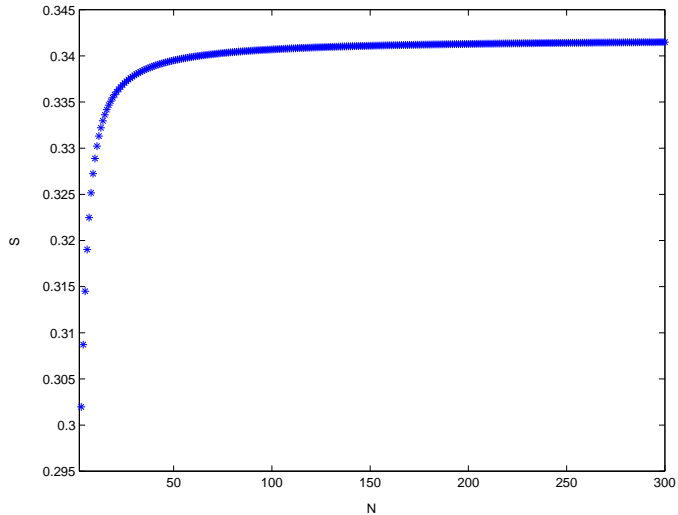}
\caption{{ The scaled entropy $S$ for a massless scalar resulting
from tracing out half of a fuzzy disc .}}
\end{center}
\end{figure}

Now, the only difference between equation (\ref{3.1}) and equation
(\ref{3.2}) is the coefficient of proportionality which can be
attributed to the shape of the boundary. When we traced out a
sub-disc ${\bf D}_n$ the boundary is a sort of a fuzzy circle with
$2n+1$ degrees of freedom, whereas tracing out the lower or upper
half of the disc the resulting boundary is linear with $N$ degrees
of freedom.

Before going into the  interpretation of our results we note that
in the fuzzy disc we have restricted the calculation to the
massless case. Indeed unlike the fuzzy sphere none  of the
matrices $W^{(m)}$ is singular, and this is due to the fact that
the fuzzy Laplacian  in the massless case has no zero modes
because of the breaking of translation invariance, therefore the
massless case does not need special treatment. Furthermore for
massive field it can easily be shown (numerically) that unless the
mass is of the order of the cutoff e.g $\mu^2\theta \sim 1$, the
resulting entropy has the same value as the massless case.
However, if the $\mu^2\theta \sim 1$ the resulting entropy differs
slightly from the massless case but starts to converge towards the
value of the massless case for large $N$.

Now  we turn to the area interpretation of our results. First it
must be noted that despite the work of \cite{liz} the fuzzy disc
is still a poor understood object and our discussion are at the
tentative and qualitative level . The radius of the fuzzy disc is
defined through the equation $R^2 =N \theta $. In order to
interpret equation (\ref{3.2}) as an area low in two dimension let
us define the effective short distance cutoff for the
disc\footnote{Indeed the entropy is dimensionless and in the free
massless theory our action does not know about the scale $\theta$
and all we  can obtain is a pure number. However the fact that the
entropy is proportional to the number of degrees of freedom on the
boundary makes the area interpretation natural. } to be $\lambda_N
=R/N= \sqrt{\frac{\theta}{N}}$. Thus equation (\ref{3.2}) becomes
( with $D= 2R$ is the diameter of the disc )
\begin{equation}\label{3.4}
  S_N=0.17 \frac{D}{\lambda_N}.
\end{equation}
In the continuum limit $N\rightarrow \infty$ with $R$ fixed ( the
limit of the continuum disc of radius $R$ ) we recover the usual
UV divergence since in that limit ${\lambda}_N{\longrightarrow}0$.

Similarly we can write equation (\ref{3.1})  ( with $r^2=n\theta
$, $\lambda_n =R/\sqrt{Nn}= \sqrt{\frac{\theta}{n}}$ and $C$ is
the circumference of the disc given by $C=2\pi r$ ) as
\begin{equation}\label{3.5}
  S_N(n)=\frac{0.23}{\pi}
\frac{C}{\lambda_n}.
\end{equation}
Again in the commutative limit  $n\rightarrow \infty$ with $r$
fixed ( the limit of the continuum disc of radius $r$ ) we recover
the usual UV divergence.

Although the two resulting entanglement entropies considered in
the fuzzy disc  case diverge in the  commutative limits of the
continuum discs they have different limits when $N \rightarrow
\infty$ with $\theta$ fixed which is the noncommutative limit of
the Moyal plane.

In the case of equation (\ref{3.5}) if we keep the size of the
small disc finite i.e $n$ finite, the result remains finite even
in the limit $N \rightarrow \infty $. Taking our calculation as a
regularization of the Moyal plane then we would expect the
entanglement entropy resulting from ignoring a finite region
(disc) to be finite and proportional to the number of degrees of
freedom on the boundary of the ignored region.

For the case of equation (\ref{3.4}) taking the limit $N
\rightarrow \infty $ amounts also to recovering the Moyal plane.
However the region which we are ignoring now is half of the Moyal
plane and it is not of finite size. From equation (\ref{3.4}) we
see that the entropy per unit length which is given by $S/D =0.17
\sqrt{\frac{N}{\theta}}$ diverges as $N$ approaches infinity. But
this divergence is not of UV origin since it shows up as a
consequence of blowing up the ignored region, i.e the divergence
is a consequence of integrating out an infinite number of degrees
of freedom. Furthermore we observe that regardless of the value of
$N$    the entropy per unit length diverges in the limit $\theta
\rightarrow 0$ which is the standard UV divergence. Indeed in the
continuum case the entropy ( per unit area ) can be rendered
finite by just introducing short distance cutoff in the normal
direction, and no IR cutoff is needed.
 Therefore the divergence of
the entropy in this Moyal plane limit  can be understood as a
UV-IR mixing rather than just coming from UV origin.

\section{Conclusions and outlook}
We have computed numerically the entanglement entropies of a
scalar field on different fuzzy spaces.  For the fuzzy sphere we
have shown that the entropy resulting from tracing out half of the
sphere is proportional to the number of non-commutative degrees of
freedom on the boundary.

The same problem was considered on a fuzzy disc and two ways of
tracing were considered. In the first case we computed the entropy
resulting from tracing out the degrees of freedom residing in a
smaller sub-disc. In the second case we considered the entropy
resulting from tracing half of the fuzzy disc. In both cases the
entanglement entropy turned out to be proportional to the number
of degrees of freedom on the boundary. In the commutative limit
both entropies suffer from the standard UV divergences. However in
the Moyal plane limit the two entropies showed different behavior.
Whereas the first entropy, resulting from ignoring the degrees of
freedom inside  a fuzzy sub-disc, remained finite when the disc
becomes a Moyal plane, the second entropy per unite area (length)
resulting from ignoring  half of the fuzzy disc diverges in the
Moyal plane limit.

The divergence in the Moyal limit could easily be seen to be
arising from  an UV-IR mixing. Therefore we would in general
expect non-commutativity to render the entanglement entropy finite
as long as the ignored region is of a finite size, as is the case
for the black hole. However, if the ignored region is not of a
finite size, the entropy per unite area may still be divergent but
now due to an IR-UV mixing. Finally it would be interesting to
investigate the divergences of the entanglement entropy per  unite
area in the Moyal plane using analytical method not just
numerical. One possible approach would still be to approximate the
Moyal plane by a fuzzy disc and use $1/N$ expansion to evaluate
the entropy.
 \paragraph{Acknowledgements}
The author D.Dou  would like to thank the Dublin Institute for
Advanced Studies for the kind hospitality during the first stage
of the work. The work of D.D is supported by MESRS, Algeria under
the research project D/39/2005 and the Associate Scheme of Abdus
Salam ICTP. The author B.Ydri would like to thank the Dublin
Institute for Advanced Studies where this research was started.
Both authors  wish to thank Denjoe O'Connor and A.Chamsa for there
extensive discussions and critical comments while this research
was in progress.

This research is dedicated to Rafael Sorkin on the occasion of his 60th birthday.
\bibliographystyle{unsrt}

\end{document}